\documentclass[11pt,a4paper]{article}
\usepackage[left=15mm, top=10mm, right=15mm, bottom=20mm, nohead=true, nofoot=true]{geometry}

\usepackage[utf8]{inputenc}
\usepackage{authblk}
\usepackage{indentfirst}
\usepackage{misccorr}
\usepackage{graphicx}
\usepackage{amsmath}
\usepackage{amssymb}
\usepackage{multicol}
\usepackage{bm}
\usepackage{longtable}
\usepackage{pdflscape}
\usepackage{epstopdf}
\usepackage{multirow}
\usepackage{adjustbox}
\usepackage{amsfonts}
\usepackage{ifthen}
\usepackage{fancyhdr}
\usepackage{setspace}
\usepackage{xcolor}
\usepackage[round,authoryear,comma,sort&compress,numbers]{natbib}
\usepackage[toc,title,page]{appendix}
\usepackage[hidelinks]{hyperref}
\usepackage{url}
\usepackage{color,soul}

\hypersetup{
    colorlinks=true,
    linkcolor=green,
    citecolor=blue,
    filecolor=magenta,
    urlcolor=cyan,
    pdftitle={Sharelatex Example},
    pdfpagemode=FullScreen,
}

\fancypagestyle{plain}{\chead{\texttt{\textcolor{brown}{Communications of Byurakan Astrophysical Observatory}}}}
\title{\textbf{The runaway nature and origin of
$\alpha$ Crucis system}}
\author[1]{M. Torosyan\thanks{marinetorosyan892@gmail.com, Corresponding author}}
\author[2]{N. Azatyan\thanks{nayazatyan@bao.sci.am}}
\author[2]{E. Nikoghosyan\thanks{elena@bao.sci.am}}
\author[2]{A. Samsonyan\thanks{anahit.sam@gmail.com}}
\author[2]{D. Andreasyan\thanks{derenik.andreasyan@gmail.com}}
\affil[1]{\scriptsize Astronomy Department, Yerevan State University, 1 Alex Manoogian Street, 0025 Yerevan, Armenia}
\affil[2]{\scriptsize Byurakan Astrophysical Observatory, 0213 Aragatsotn Province, Armenia}

\def\apj{Astrophys.~J. }

\def\aap{Astron.~Astrophys. }

\def\apjl{Astrophys.~J.~Lett. }

\def\aj{Astron.~J. }
\def\mnras{Mon.~Not.~R.~Astron.~Soc. }
\def\araa{Ann.~Rev.~Astron.~Astrophys. }

\def\bain{Bull.~Astron.~Inst.~Neth. }

\begin{document}
\pagestyle{empty}
\newpage
\pagestyle{fancy}
\label{firstpage}
\date{}
\maketitle
\begin{abstract}
Massive stars are always the focus of astronomical research and a significant part of them (10--20\%) moves in space at a high (supersonic) velocity. This paper presents the results of a study of the $\alpha$ Crucis system, located at $\sim$114\,pc distance from the Sun, with an observed bow shock around it. We used data and images from the Gaia and WISE space telescopes. The coordinates, distance, and proper motion of the $\alpha$ Crucis system were used to determine its space velocity. We managed to find a stellar cluster to which the $\alpha$ Crucis system belongs, that is, it has not been ejected from its parent cluster, but is moving in space together with other members of the cluster. The $\alpha$ Crucis system has a velocity of $\sim$1.3\,km/s relative to the star cluster. The geometric parameters of the bow shock are compatible with other known bow shocks. The bow shock is unaligned, i.e., most likely interstellar medium (ISM) large-scale motions are responsible for the resulting bow shock, which is further evidence that the $\alpha$ Crucis system is not runaway in nature.
\end{abstract}

\section{Introduction}

The massive stars within the Milky Way are not randomly dispersed; approximately 80\% of the O-type stars are part of OB associations located in the galaxy's spiral arms \citep{Brown1999}. The kinematic characteristics of the remaining 20\%, which belong to the field population, indicate that these O stars are runaways—originating from OB associations but subsequently escaping \citep{Blaauw1993}. Two primary scenarios are widely accepted to explain the presence and high velocities of runaway stars: (1) the supernova explosion of a companion star within a massive binary system \citep{Blaauw1961}, and (2) the dynamic ejection from a young star cluster during its early developmental stages \citep{Poveda1967}. Identifying the original OB association of a runaway star is crucial, as it provides insights into the star's evolutionary history. By knowing the proper motions of both the runaway star and its parent OB association, the kinematic age of the runaway star can be determined.

Another method of searching for high-velocity runaway star candidates is to bow-shocks associated with these objects, which are generated by the interaction of the star's wind with the interstellar medium (ISM) \citep[e.g., ][Azatyan et al. submited]{Buren1988,Peri2012}. Therefore, OB runaways can be used also as probes of the  ISM. As rule,  bow shocks emit strong radiation in the mid-infrared range \citep[e.g., ][]{Gvaramadze2011}.

\begin{figure}[h!]
    \centering
    \includegraphics[width=0.6\textwidth]{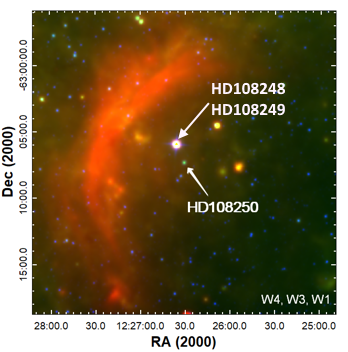}
    \caption{Colour-composite image of $\alpha$ Crucis system and the bow shock around it: 3.4\,$\mu$m (blue), 12\,$\mu$m (green), and 22\,$\mu$m (red) bands. The three stars of the system are indicated by white arrows.}
    \label{fig:system}
\end{figure}

In our paper, we have chosen the $\alpha$ Crucis system as the target of study. The $\alpha$ Crucis system consists of three stars: HD\,108248, HD\,108249, and HD\,108250. HD\,108248 and HD\,108249 are known as a$^1$ Cru and a$^2$ Cru with spectral types B0.5IV and B1V, respectively \citep{Houk1975,Reed2005}. HD\,108250 is also known as $\alpha$ Cru C with spectral type B4V \citep{Tokovinin1999, Reed2005}. Figure \ref{fig:system} shows a combination of 22\,$\mu$m  (red), 12\,$\mu$m (green), and 3.4\,$\mu$m (blue) images of the $\alpha$ Crucis system and the bow shock detected around it. The three stars of the system are indicated by white arrows. The HD\,108248 and HD\,108249 pair is significantly brighter compared to HD\,108250. HD\,108250 is 90 arcseconds away from the pair HD\,108248 and HD\,108249 and has the same space motion, suggesting they may be gravitationally bound \citep{Shatsky2002}. \citet{Rizzuto2011} calculated that this system, including HD\,108250, is a member of the Lower Centaurus--Crux subgroup in the Scorpius--Centaurus association with a 66\% probability. The radial velocities of stars HD\,108248, HD\,108249, and HD\,108250 are 11.2$\pm$2, 0.6$\pm$2, and 27$\pm$5\,km/s, respectively \citep{Wilson1953}. The parallaxes and proper motions of the three stars in the system are almost the same. The parallaxes and proper motions of HD\,108248 and HD\,108249 are 10.13$\pm$0.50\,mas, -35.83$\pm$0.47\,mas/yr, and -14.86$\pm$0.43\,mas/yr, respectively \citep{Leeuwen2007}. The third data release of Gaia observatory (Gaia\,DR3)  provides reliable information only on the star HD\,108250. The parallax is 9.3964$\pm$0.1289\,mas \citep{Gaia2022}, giving a distance of $\sim$114\,pc from the Sun. The fractional parallax uncertainty (\texttt{parallax/parallax\_error}) is $\sim$73. The proper motion in RA is -39.548$\pm$0.117\,mas/yr and in Dec is -13.760$\pm$0.132\,mas/yr \citep{Gaia2022}. The Renormalized UnitWeight Error (\texttt{ruwe}) coefficient is 1.316, indicating good quality astrometric solutions for this star.

Although the star HD\,108250 is at the center of symmetry of the observed bow shock, this bow shock has previously been associated with the HD\,108248 and HD\,108249 pair \citep{Bodensteiner2018}. Since these three stars are physically connected, the observed bow shock is most likely associated with all three. The aim of this work is to elucidate the runaway nature of the $\alpha$ Crucis system, find the parent star cluster, calculate the relative velocity of the system with respect to the parent star cluster, and determine the kinematic age. Another goal is to study the properties of the observed bow shock.

We have organised the paper as follows: Section \ref{data} describes the data used; Section \ref{Res} details the results along with discussions, covering the search for a parent star cluster, the determination of the relative velocity of the $\alpha$ Crucis system, and the characteristics of the observed bow shock; the summarising of the results is in Section \ref{Conclusion}.

\section{Data used}
\label{data}
Since the bow shock observed around the $\alpha$ Crucis system radiates in the mid-infrared range, we used the Wide-field Infrared Survey Explorer \citep[WISE;][]{Wright2010} images at 3.4, 4.6, 12, and 22\,$\mu$m wavelengths. For kinematic parameters, we used the Gaia Data Release 3 \citep[Gaia DR3;][]{Gaia2022} catalog data.

\section{Results and discussion}
\label{Res}
\subsection{Searching for a parent star cluster}
\label{parent}

\begin{figure}[h!]
    \centering
    \includegraphics[width=0.480\textwidth]{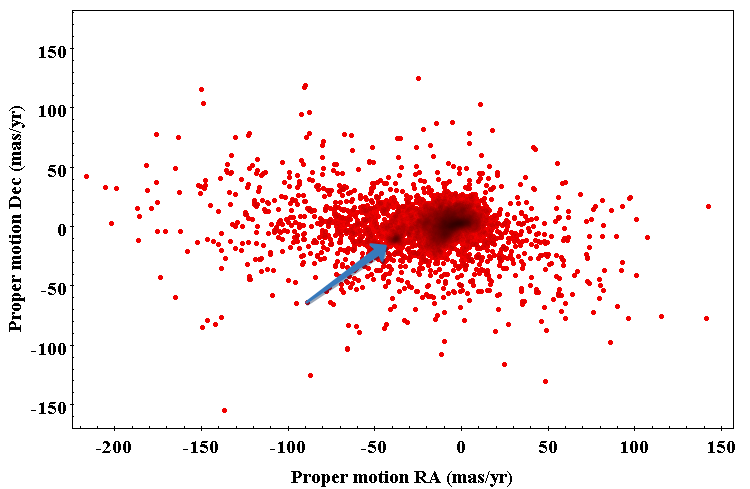}
    \includegraphics[width=0.405\textwidth]{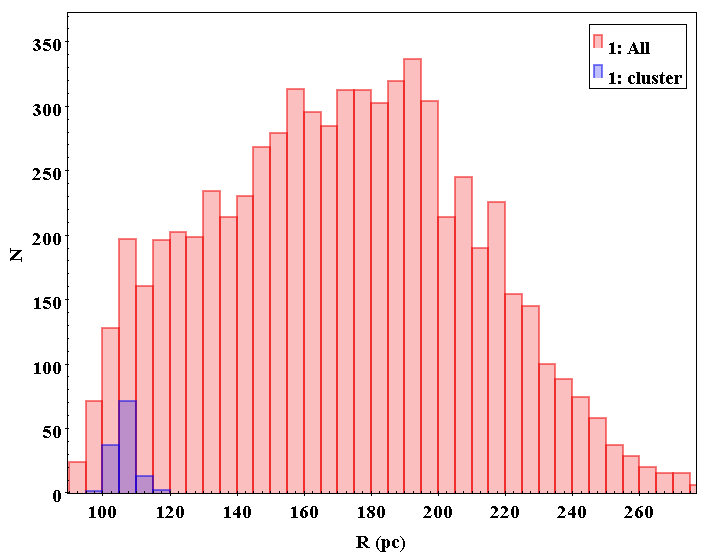}
    \caption{Left: Distribution of proper motions of the objects obtained based on the search in the direction of the star HD\,108250. Red circles are objects obtained based on the search. Blue arrow shows the star cluster clearly separated from the field. Right: Histogram of object distances. The red histogram is for all objects, and the blue one is for the cluster members.}
    \label{fig:fig2}
\end{figure}

We searched for members of candidate parent stellar cluster/association of HD\,108250 in the Gaia\,DR3 catalog in the range $(l, b) = (291.0^\circ \text{ to } 320.0^\circ, -3.5^\circ \text{ to } +3.5^\circ)$. Since Gaia\,DR3 only provides reliable data for the star HD\,108250, we assume that the obtained results are correct for the entire system. We apply a series of filters on the data, which are briefly described below:
\begin{itemize}
    \item We excluded sources with \texttt{ruwe} greater than 1.4, which probably indicates poor astrometric solutions for these sources.
    \item We excluded sources for which the ratio of parallax to parallax uncertainty (\texttt{parallax/parallax\_error}) was less than 5.
    \item We excluded sources for which the parallax is not in the range of 8.3 to 10.3\,mas. 
\end{itemize}
As a result, 3959 sources were obtained. The obtained sample were reflected by Tool for OPerations on Catalogs And Table \citep[TOPCAT;][]{Possel2020}  through the distribution of proper motions. The left panel of Figure \ref{fig:fig2} shows the distribution of the proper motions of objects obtained based on the search in the direction of the star HD\,108250. The blue arrow indicates a star cluster with its proper motions clearly separated from the  field. The star HD\,108250, with its measured proper motions and distance, most likely belongs to that cluster. On the right panel of Figure \ref{fig:fig2} is the histogram of object distances. The red one is the histogram of all objects, and the blue one -- the cluster members, which again stands out clearly from the field stars and has a maximum at the distance where the $\alpha$ Crucis system is located. There are 93 members of the cluster. The average parallax of the cluster members is $9.33\pm0.25$\,mas, which means that the cluster is located at $107\pm3$\,pc  distance. The average proper motion of members in the direction of RA is $-37.91\pm0.89$\,mas/yr, and in the direction of Dec is $-10.62\pm1.41$\,mas/yr. Performing a comparison between the parameters of the star HD\,108250 and the members of the cluster, namely the parallax (and hence the distance) and proper motions,  it is clear that the star HD\,108250 or the $\alpha$ Crucis system has not escaped from its parent cluster and is gravitationally bound to the other members, i.e., together they are moving in space with almost the same space velocity. The cluster is one of the subgroups in the Scorpius--Centaurus association, and these subgroups are located quite far from each other, but move as a whole \citep{Blaauw1964, Goldman2018}.

\begin{figure}[h!]
    \centering
    \includegraphics[width=0.6\textwidth]{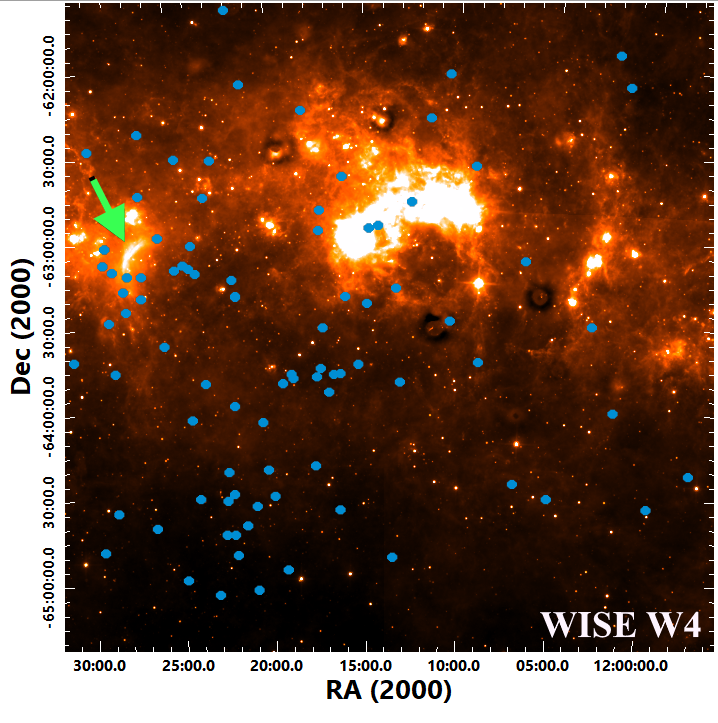}
    \caption{Distribution of the cluster members in the field on WISE W4 (22\,$\mu$m) image. The members are marked with blue circles. The green arrow shows the position of the bow shock observed around the $\alpha$ Crucis system.}
    \label{fig:fig3}
\end{figure}

Figure \ref{fig:fig3} shows the distribution of cluster members in the field. Members are marked with blue circles. The green arrow shows the position of the bow shock observed around the $\alpha$ Crucis system. As can be seen, the members have a large dispersion in the field and the $\alpha$ Crucis system is located in the peripheral part of the star cluster.

\subsection{Relative velocity of the $\alpha$ Crucis system}
\label{Velocity}

One of the conditions for the runaway nature of the $\alpha$ Crucis system is its high space velocity relative to its parent star cluster. To determine the space or peculiar velocity relative to the parent cluster, the coordinates, distance, proper motions, and apparent radial velocities of HD\,108250 and the parent cluster members are required. Since the radial velocities in Gaia\,DR3 are known for only a part of the cluster members, i.e., the data are incomplete, we therefore set the radial velocity component equal to zero and calculated the relative transverse velocity of HD\,108250. Therefore, the  peculiar velocity of the star HD\,108250 or the $\alpha$ Crucis system relative to the parent cluster is approximately 1.3\,km/s. This finding supports the conclusion that the $\alpha$ Crucis system, despite the bow shock observed around it, is not a runaway system, since the relative velocity of runaway stars exceeds 30\,km/s \citep{Blaauw1961}.

\subsection{Bow shocks}
\label{Shock}

\begin{figure}[h!]
    \centering
    \includegraphics[width=0.6\textwidth]{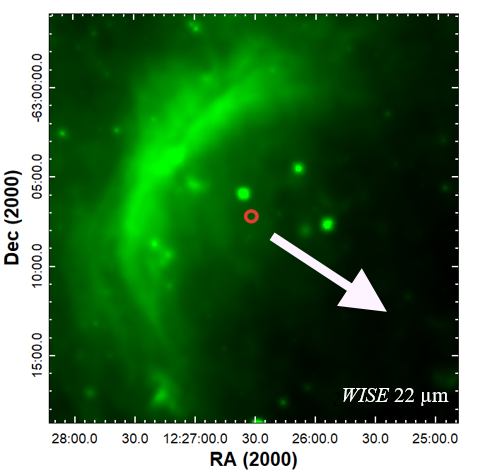}
    \caption{The bow shock observed around the $\alpha$ Crucis system on WISE W4 (22\,$\mu$m) image. Open red circle shows the location of HD\,108250 star. White arrow indicates direction of motion of the $\alpha$ Crucis system.}
    \label{fig:shock}
\end{figure}

Runaway stars move in the ISM at velocities that exceed the speed of sound. Their stellar wind hits the ISM and produces so-called bow shocks \citep{Wilkin1996}. Due to the stellar wind, the dust is heated and re-radiates at infrared wavelengths. Bow shocks have also been observed in the optical range \citep[e.g.,][]{Brown2005} and in a few cases at radio wavelengths \citep{Benaglia2010}. The $\alpha$ Crucis system and the surrounding bow shock is well resolved in the mid-infrared wavelength range. Figure \ref{fig:shock} shows the bow shock around the system in the WISE 22\,$\mu$m channel. The open red circle shows the position of HD\,108250, and the white arrow shows the direction of motion of the $\alpha$ Crucis system. As we see, the observed bow shock is unaligned, that is, the apparent motion of the system does not coincide with the angular opening of the nebula (bow shock). This can be explained by large-scale motions in the ISM, which may also be responsible for the high relative velocity between the star and the ISM \citep{Bodensteiner2018}. This is probably why the bright pair HD\,108248 and HD\,108249 in the $\alpha$ Crucis system, with which the observed bow shock was previously associated, is offset from the center of symmetry of the bow shock, which is usually observed in unaligned bow shocks \citep{Wilkin2000}. The fact that the bow shock is unaligned probably proves that the $\alpha$ Crucis system is not runaway.

To compare the properties of the observed bow shock with other known bow shocks, we measured its geometrical parameters, such as the spatial length $l = 17.2 \pm 0.57$\,pc, the width $w = 4.2 \pm 0.14$\,pc, and the distance from the star to the center of the bow shock (standoff distance) $R = 5.5 \pm 0.18$\,pc. Table 4 presented in \citet{Peri2012} paper contains the same parameters for a number of other bow shocks. We compared them with our results and concluded that the parameters of the bow shock surrounding the $\alpha$ Crucis system are compatible with the parameters of those already known bow shocks.

\section{Conclusion}
\label{Conclusion}

Within the framework of this paper, the results of the study of the $\alpha$ Crucis system at a distance of $\sim$114\,pc and the bow shock observed around it are presented. We have discovered a star cluster that is clearly separated from the stars of the field by their proper motions and distances, the number of members of which is 93. The cluster is one of the subgroups in the Scorpius-Centaurus {association. The $\alpha$ Crucis system, with its proper motion and distance, most likely belongs to that cluster. We concluded that the $\alpha$ Crucis system has not escaped from its parent star cluster and is gravitationally bound to the other members, that is, they move together in space, having almost the same space velocity. The peculiar velocity of the $\alpha$ Crucis system with respect to the star cluster is $\sim$1.3\,km/s.

We also studied the parameters of the bow shock observed in the mid-infrared wavelength around the $\alpha$ Crucis system. By measuring the geometrical parameters of this bow shock, we concluded that its properties are similar to other known bow shocks. The apparent motion of the $\alpha$ Crucis system does not match the angular opening of the nebula 
(bow shock), i.e., the observed bow shock is unaligned. This means that the bow shock was formed because of the large-scale motions in the ISM, not the velocity of the $\alpha$ Crucis system. 

Taking the above results into account, we note as the main conclusion that the $\alpha$ Crucis system, despite the bow shock observed around it, is not runaway.


\section*{\small Acknowledgements}
\scriptsize{This work was partially supported by a research grant number N\textsuperscript{\underline{o}}\,21AG-1C044 from Higher Education and Science Committee of Ministry of Education, Science, Culture and Sport RA. This work presents results from the European Space Agency (ESA) space mission Gaia. Gaia data are being processed by the Gaia Data Processing and Analysis Consortium (DPAC). Funding for the DPAC is provided by national institutions, in particular the institutions participating in the Gaia MultiLateral Agreement (MLA). This publication also makes use of data products from the Wide-field Infrared Survey Explorer, which is a joint project of the University of California, Los Angeles, and the Jet Propulsion Laboratory/California Institute of Technology, funded by the National Aeronautics and Space Administration.}

\scriptsize
\bibliographystyle{ComBAO}
\nocite{*}
\bibliography{references}

\end{document}